\documentclass[
  aps,prd,
  twocolumn,
  superscriptaddress,
  nofootinbib,
  longbibliography,
  amsmath,amssymb
]{revtex4-2}

\usepackage{graphicx}
\usepackage{microtype}
\usepackage{siunitx}
\usepackage[colorlinks=true,allcolors=blue]{hyperref}
\usepackage{bm}
\usepackage{mathtools}
\usepackage{multirow}

\usepackage{placeins} 
\usepackage[table]{xcolor} 
\usepackage{colortbl} 
\usepackage{float}
\usepackage{caption}
\usepackage{textcomp} 

\begin{document}

\title{A Novel Implementation of the Matrix Element Method at Next-to-Leading Order\\
  for the Measurement of the Higgs Self-Coupling $\lambda_{3H}$}

\author{Matthias Tartarin}
\email{matthias.tartarin@l2it.in2p3.fr}
\affiliation{Université de Toulouse, CNRS/IN2P3, L2IT, Toulouse, France}

\author{Jan Stark}
\email{jan.stark@l2it.in2p3.fr}
\affiliation{Université de Toulouse, CNRS/IN2P3, L2IT, Toulouse, France}

\date{\today}

\begin{abstract}
The determination of the Higgs boson trilinear self-coupling
$\lambda_{3H}$ is a key goal of the LHC physics programme. Its precise
measurement will provide unique insight into the scalar potential and
the mechanism of electroweak symmetry breaking. Higgs boson pair
production in the $gg\to HH$ process, and particularly in the $HH\to b\bar{b}\gamma\gamma$
final state, offers direct sensitivity to $\lambda_{3H}$.
\\
We present the first implementation of the Matrix Element Method at
Next-to-Leading Order (MEM@NLO) for this process, which is publicly available. 
The MEM is a statistically optimal approach that maximises information extraction
from collision events. Extending it to NLO represents a major
methodological challenge, which we address with a new formalism
integrated into the \textsc{MoMEMta} framework. Results with simulated
pseudo-experiments demonstrate, in a proof-of-principle study, the strong discriminating power of the method 
and its ability to extract the coupling modifier $\kappa_\lambda = \lambda_{3H}/\lambda^{SM}_{3H}$
with high precision.
\end{abstract}


\maketitle

\section{Introduction}

The discovery of the Higgs boson in 2012~\cite{ATLAS_Higgs_2012,CMS_Higgs_2012}
at the Large Hadron Collider (LHC) by the ATLAS~\cite{ATLAS_ref} and
CMS~\cite{CMS_ref} collaborations completed the experimental
confirmation of the Standard Model (SM) particle content.
Despite this milestone, the form of the Higgs potential itself remains only
weakly constrained experimentally.
\\
As the leading self-interaction of the Higgs sector, the trilinear Higgs self-coupling $\lambda_{3H}$
is particularly sensitive to modifications of the Higgs potential, and
therefore provides a powerful probe of the Higgs sector, and possible
new dynamics beyond the Standard Model.
\\
\\
Such deviations can affect both the total rate and the event-level
kinematics of Higgs boson pair production, which makes $\lambda_{3H}$
a key target of the LHC physics programme and its future phases 
(such as the High-Luminosity LHC, HL-LHC)~\cite{Future_1,Future_2}.
In a broader class of scenarios, such modifications may also play a crucial role in explaining the observed matter-antimatter
asymmetry of the Universe~\cite{baryogenesis}.
\\
\\
Despite extensive searches, di-Higgs production has not yet been observed at the LHC. 
The current limits on $\lambda_{3H}$ (often expressed via the coupling modifier 
$\kappa_{\lambda} = \lambda_{3H}/\lambda_{3H}^{\mathrm{SM}}$) 
still allow sizeable deviations from the SM prediction. 
Using the full Run~2 dataset, ATLAS and CMS have performed combined
interpretations of the major di-Higgs channels 
(with final states $b\bar{b}b\bar{b}$, $b\bar{b}\tau^+\tau^-$, $b\bar{b}\gamma\gamma$, 
and multileptons), 
yielding constraints at 95\% confidence level of
$-1.2 < \kappa_\lambda < 7.2$ from ATLAS~\cite{ATLAS_kappa}
and $-1.4 < \kappa_\lambda < 7.0$ from CMS~\cite{CMS_kappa}.
\\
The most recent ATLAS analysis in the $b\bar{b}\gamma\gamma$ final state,
combining Run~2 and early Run~3 data 
(corresponding to a total integrated luminosity of 308~fb$^{-1}$), sets limits on the coupling modifier 
$-1.4 < \kappa_\lambda < 6.8$ at 95\%~CL~\cite{bbyy_Analysis_2025}, 
representing the most stringent constraints obtained to date in this
channel. 
While increased luminosity is a necessary ingredient to improve statistical precision 
on measurements, it is not sufficient alone. Fully exploiting the available data also 
requires analysis techniques capable of extracting the maximal discriminating power 
from event-level kinematics.
\\
\\
The Matrix Element Method (MEM)~\cite{MEM_0,MEM_0bis} provides a natural
framework to address this challenge.
By exploiting the full event-level kinematic information through
first-principles likelihoods, it is particularly well suited to the
extraction of parameters associated with rare processes.
The MEM has been successfully applied to measurements with limited statistics and significant background, 
most notably for the determination of the top quark mass at the Tevatron~\cite{MEM_1,
MEM_2, MEM_3}. 
\\
\\
The measurement of Higgs boson pair production at the LHC presents
closely analogous challenges, as the extraction of the Higgs
self-coupling $\lambda_{3H}$ relies on a process with very small production rates and
limited statistical sensitivity.
However, most existing applications of the MEM rely on Leading-Order (LO) matrix
elements, neglecting higher-order QCD radiation. As a result, LO-based MEM analyses suffer from reduced accuracy 
when applied to realistic LHC events, motivating the development of strategies to incorporate
next-to-leading order (NLO) QCD effects within MEM-based frameworks
\cite{1010.2263,1204.4424,1712.04527,2301.03280}.
\\
\\
The strategy presented in this work introduces a new framework and realises 
the first MEM at Next-to-Leading Order (NLO)~\cite{Tartarin_2025} for the process 
$gg \to HH \to b\bar b \gamma\gamma$, enabling a precise and robust extraction
of $\kappa_{\lambda}$ for realistic events.
This study follows up on the results at LO only, reported in Ref.~\cite{Feble}.

\newpage
\section{The Matrix Element Method}

The Matrix Element Method (MEM) provides a statistically optimal way to
exploit the full kinematic information from collision events, by evaluating
the likelihood that a measured event with reconstructed observables
$\mathbf{x}^i$ originates from a given process $p$ under a set of model
hypotheses~$\mathbf{h}$.
\\
\\
Each process $p$ corresponds to a specific hard-scattering topology, 
and is characterised by parameters
$\mathbf{h}$ such as the coupling modifier
$\kappa_{\lambda}$. 
The relevant signal process for this analysis is
gluon-fusion ($ggF$) di-Higgs production $gg \to HH \to b\bar{b}\gamma\gamma$,
while the main background processes include $t\bar{t}H$,
QCD-induced diphoton production, and single-Higgs (plus-jets) production.
\\
\\
For a given event $i$, the reconstructed quantities $\mathbf{x}^i$
denote the four-momenta of the observed final-state objects.
The MEM then integrates over all parton-level four-momenta
$\mathbf{y}$ that are compatible with these reconstructed measurements.
From first principles, the corresponding process likelihood for event
$i$ can be expressed as:

\begin{align}
\mathcal{L}_{\text{process}}^{\text{p}}(\mathbf{h}|\mathbf{x}^i)
 &= \frac{(2\pi)^4}{\sigma^{\text{obs}}_p(pp\!\to\!F)}
    \int_{\mathbf{y}}\!
    \int_{q_1,q_2}
    \sum_{a_1,a_2}
    f_{a_1}(q_1) f_{a_2}(q_2)\, \nonumber\\
  &\times 
    \frac{\bigl|\mathcal{M}_p(a_1 a_2 \!\to\! \mathbf{y}; \mathbf{h})\bigr|^2}{q_1 q_2 s}
    W(\mathbf{y}, \mathbf{x}^i)\ \nonumber\\
  &\times 
    \delta\!\Bigl(a_1 + a_2 - \sum_{j=1}^{n} y_j\Bigr)\,
    dq_1\,dq_2\,d^{4n}\mathbf{y},
  \label{eq:likelihood}
\end{align}

\noindent
where the various quantities entering Eq.~\eqref{eq:likelihood} are:
\begin{itemize}
  \item $\sigma^{\rm obs}_p(pp\!\to\!F)$ which corresponds to the observable 
        integrated cross section (\textit{i.e.} the fiducial cross section multiplied by the detector efficiency) to produce the
        final-state partons $F$ for the process $p$, used to normalise the likelihood;
  \item $q_1$ and $q_2$ are the momentum fractions of partons $a_1$ and $a_2$ (also referred to as Bjorken-$x$ scaling variables),
        and $s$ is the Mandelstam variable corresponding to the squared centre-of-mass 
        energy of the proton-proton collision;
  \item $\mathcal M_p(a_1 a_2\!\to\!\mathbf y;\mathbf h)$
        denotes the partonic matrix element of
        $a_1 a_2\!\to\!\mathbf y$ for the process $p$ and the hypothesis~$\mathbf h$, 
        obtained from first principles in quantum field theory;
  \item $\delta^{(4)}\!\bigl(a_1{+}a_2{-}\sum_j y_j\bigr)$ are $\delta$-functions that enforce 
        energy-momentum conservation;
  \item $W(\mathbf{y},\mathbf{x}^i)$ corresponds to the transfer functions, which
          parameterise the detector response and describe the probability for a parton-level configuration 
          $\mathbf{y}$ to give rise to the reconstructed observables $\mathbf{x}^i$.
\end{itemize}

\vspace{0.5em}
\noindent
\textsc{MoMEMta}~\cite{MoMEMta} is a modular framework designed to implement the
MEM in high-energy physics analyses through numerical
evaluation of the integral in Eq.~\eqref{eq:likelihood}. It provides a flexible environment in which
the ingredients of the MEM integrand are defined and
handled as separate \emph{modules}.
Yet, because the MEM involves integrations over a high-dimensional phase space, 
the numerical complexity can quickly become
very challenging. 
To make these calculations more manageable, \textsc{MoMEMta} uses 
dedicated modules called \emph{blocks} that perform analytic changes
of variables using kinematic constraints 
(\textit{e.g.} from intermediate resonances such as the Higgs boson) 
and conservation laws encoded in the $\delta$-functions to reduce
the dimensionality of the integration.
\\
\\
At LO, the \textsc{MoMEMta} modular approach
performs remarkably well, as the phase space has a simple structure and the
mapping between partonic and reconstructed objects is straightforward. 
\\
\\
At NLO however, additional complications arise. While virtual corrections preserve
the LO final-state topology, real-emission contributions introduce an extra
parton and therefore a different phase space configuration. 
For this analysis, the
extra radiation will be treated as unresolved 
(\textit{i.e.} no attempt is made to reconstruct the additional parton in the detector) 
and will be fully integrated over. 
This allows all NLO contributions to be consistently mapped onto the LO-like final-state $\mathbf{x}^i$,
corresponding to the kinematics of reconstructed $b\bar{b}\gamma\gamma$ event.

\section{Accessing the matrix elements at NLO}

At NLO, the differential cross section separates into three
contributions:
\begin{equation}
  \mathrm{d}\sigma_{\text{NLO}}
  = \underbrace{\bigl[B(\Phi_n)+V(\Phi_n)\bigr]\,\mathrm{d}\Phi_n}_{\text{Born+Virtual}}
  \;+\;
    \underbrace{R(\Phi_{n+1})\,\mathrm{d}\Phi_{n+1}}_{\text{Real}},
  \label{eq:nlo-decomp}
\end{equation}
where $B$ denotes the Born (or LO) contribution, $V$ the virtual loop correction with
the same final-state multiplicity, and $R$ the real-emission matrix element
with one additional parton. 
\\
\\
The cancellation of soft and collinear singularities between the virtual
and real contributions must be treated consistently to ensure that the
separate Born+Virtual and Real-emission terms remain finite and
numerically stable.
As discussed in the following subsections, our framework interfaces the
MEM integrand directly with these finite, regularised NLO matrix element
contributions on a point-by-point basis, enabling the explicit
construction of the likelihood while preserving infrared safety and
numerical stability across the entire phase space.

\newpage
\noindent\textbf{Virtual vs.\ Real: practical challenges.}\\
As one can see from Eq.~\eqref{eq:nlo-decomp}, virtual contributions share 
the same kinematic topology as the Born term
and can, in principle, be evaluated over the same $\mathrm{d}\Phi_n$
phase space. However, they contain infrared poles and depend on specific conventions and must be
handled consistently to ensure numerical stability and proper matching
with the real-emission component.
\\
\\
Real emission, in contrast,
lives in the $\mathrm{d}\Phi_{n+1}$ phase-space and introduces an extra parton and integration
variables. In the MEM framework, this additional radiation does not need to be
explicitly reconstructed at detector level. Instead, it can be consistently accounted 
for by integrating over its corresponding extra three degrees of freedom.
This requires a careful treatment of the soft and collinear regions to
maintain infrared safety and numerical stability in the MEM.

\vspace{0.5em}
\noindent\textbf{Strategy: repurposing \textsc{POWHEG-BOX-V2} to access pointwise matrix elements.}\\
Computing the likelihood $\mathcal{L}_{\text{process}}^{\text{p}}(\mathbf{h}|\mathbf{x}^i)$ 
requires access to the matrix element
$\mathcal{M}_p$, evaluated at specific phase-space points. 
At NLO, this involves going beyond the leading
Born term by including both virtual loop corrections and 
real-emission contributions.  
\\
\\
To obtain pointwise access to the amplitudes at arbitrary phase-space
points, we use the \textsc{POWHEG-BOX-V2} software~\cite{POWHEG_MANUAL}, and in
particular its dedicated ggHH subrepository developed by
Heinrich \textit{et al.}~\cite{Heinrich_1,Heinrich_2}. This implementation
provides configurable $\kappa_\lambda$ and top-mass schemes.  
\\
\textsc{POWHEG-BOX-V2} is a general NLO Monte Carlo framework designed
for the computation of cross sections and the generation of fully
exclusive events at NLO accuracy, with subsequent parton-showering 
provided via its interface to \textsc{PYTHIA}~\cite{pythia82}. 
\\
\\
By construction, \textsc{POWHEG-BOX-V2} does not
natively provide external access to the individual Born, Virtual, and Real contributions at arbitrary phase-space points. 
A central part of this work therefore consisted in adapting the
internal routines of \textsc{POWHEG-BOX-V2} and developing a dedicated
interface to \textsc{MoMEMta}, enabling point-by-point access to these
amplitudes for user-specified phase-space configurations.
Accessing each of them is required for the explicit evaluation of the matrix element entering the likelihood defined in
Eq.~\eqref{eq:likelihood}.
\\
This interface between the two softwares is a key component of this new MEM@NLO implementation, 
and a detailed description of its structure and the validation
procedures are given in Ref.~\cite{Tartarin_2025}.
\\
\\
To promote open access and ensure reproducibility of our results, a
public version of this study, together
with additional auxiliary material, is provided in
Ref.~\cite{Tartarin_phd_share}.


\section{Interfacing to \textsc{MoMEMta}.}
\label{sec:inter_mom}

As mentioned earlier, real-emission contributions at NLO introduce an
additional parton in the final state, modifying both the dimensionality
and structure of the phase-space integration, even when the emitted
radiation is treated as unresolved.
\\
The \textsc{MoMEMta} framework, originally designed for LO
implementations of the MEM, does not include any block
capable of handling the integration of real NLO contributions.  
To overcome this limitation, we developed a new component,
\textsc{Block~N}.

\begin{figure}[h]
  \centering
  \includegraphics[width=1.1\columnwidth]{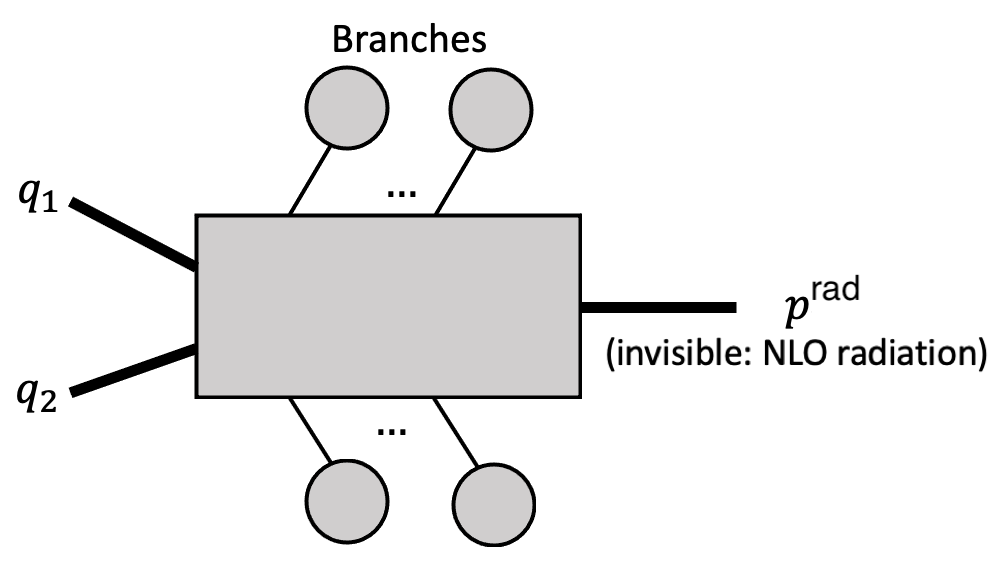}
  \caption{Schematic representation of the new \textsc{Block~N}, following the conventions used by \textsc{MoMEMta}.}
  \label{fig:blockn}
\end{figure}

\vspace{0.8em}
\noindent\textbf{New \textsc{Block~N}: Design and principle}\\
Blocks in \textsc{MoMEMta} are used to reduce the dimensionality of
the integral in Eq.~\eqref{eq:likelihood} through well-chosen changes of
variables.
Building on a systematic study of all existing blocks and their Jacobians, 
we developed a new block that we called \textsc{Block~N}, 
specifically designed to incorporate the additional
degrees of freedom related to the unresolved real emissions at NLO.
\\
\\
A graphical representation of this new \textsc{Block~N} is shown in Fig.~\ref{fig:blockn}.

\noindent
The design principle of the new \textsc{Block~N} is to simultaneously eliminate the two Bjorken-$x$
fractions ($q_1$, $q_2$) and the transverse components of the radiated
parton momentum ($p_x^{\text{rad}}$, $p_y^{\text{rad}}$) using 
four-momentum conservation, while keeping its longitudinal
component $p_z^{\text{rad}}$ as a free integration variable.  
\\
\\
The choice to keep $p_z^{\text{rad}}$ is also motivated by physical considerations: 
its direction coincides with the unknown net boost of the system of the two initial-state partons, 
along the beam axis, which cannot be reconstructed from the final state alone.  

\newpage
\noindent
The corresponding Jacobian associated to the \textsc{Block~N} is derived from the system:  
\[
\begin{cases}
p^{\text{rad}}_{x} = -P_x,\\
p^{\text{rad}}_{y} = -P_y,\\
q_{1,E} - q_{2,E} = P_z,\\
q_{1,E} + q_{2,E} - \sqrt{(p^{\text{rad}}_{x})^2+(p^{\text{rad}}_{y})^2+(p^{\text{rad}}_{z})^2}=E_T,
\end{cases}
\]

\noindent
where $P_x$, $P_y$, $P_z$ and $E_T$ are the sums of the corresponding
contributions over final-state $b\bar{b}\gamma\gamma$,
$p^{\text{rad}}$ denotes the unresolved real-radiation, and
$q_{i,E} = \frac{1}{\sqrt{2}} q_i \sqrt{s}$ is the energy of gluon $i$.
\\
\\
From this system of equations, we can compute the Jacobian associated with the change of variables:

\begin{equation}
  \begin{aligned}
    J^{-1} 
    &=
    \begin{pmatrix}
    \frac{\partial P_x}{\partial p^{\text{rad}}_{x}} & \frac{\partial P_x}{\partial p^{\text{rad}}_{y}} & \frac{\partial P_x}{\partial q_{1,E}} & \frac{\partial P_x}{\partial q_{2,E}} \\
    \frac{\partial P_y}{\partial p^{\text{rad}}_{x}} & \frac{\partial P_y}{\partial p^{\text{rad}}_{y}} & \frac{\partial P_y}{\partial q_{1,E}} & \frac{\partial P_y}{\partial q_{2,E}} \\
    \frac{\partial P_z}{\partial p^{\text{rad}}_{x}} & \frac{\partial P_z}{\partial p^{\text{rad}}_{y}} & \frac{\partial P_z}{\partial q_{1,E}} & \frac{\partial P_z}{\partial q_{2,E}} \\
    \frac{\partial E_T}{\partial p^{\text{rad}}_{x}} & \frac{\partial E_T}{\partial p^{\text{rad}}_{y}} & \frac{\partial E_T}{\partial q_{1,E}} & \frac{\partial E_T}{\partial q_{2,E}}
    \end{pmatrix}
    \\[1.2ex]
    &=
    \begin{pmatrix}
    -1 & 0 & 0 & 0 \\
    0 & -1 & 0 & 0 \\
    0 & 0 & 1 & -1 \\
    -\dfrac{p^{\text{rad}}_{x}}{E^{\text{rad}}} & -\dfrac{p^{\text{rad}}_{y}}{E^{\text{rad}}} & 1 & 1
    \end{pmatrix} .
    \end{aligned}
    \label{eq:detMatrix}
\end{equation}

\noindent
The determinant of the inverse Jacobian matrix $J^{-1}$ reads:

\begin{equation}
  \begin{aligned}
    \left| \det(J^{-1}) \right|
    &= \left| (-1)(-1)
        \cdot
        \det \begin{pmatrix}
          1 & -1 \\
          1 & 1
        \end{pmatrix} \right| \\
    &= \left| \det \begin{pmatrix}
          1 & -1 \\
          1 & 1
        \end{pmatrix} \right|
       = |2| = 2 ,
  \end{aligned}
  \label{eq:detJInv}
\end{equation}

\noindent
and the determinant that needs to be implemented in the code for \textsc{Block~N} is:
\begin{equation}
    \left| \det(J) \right| = \frac{1}{\left| \det(J^{-1}) \right|} = \frac{1}{2}.
    \label{eq:detJ}
\end{equation}

\vspace{0.5em}
\noindent\textbf{Validation of \textsc{Block~N}}\\
Given the novelty of the Block~N construction, a dedicated validation
strategy was required to ensure the correctness of the phase-space
transformation and its associated Jacobian.
\\
A detailed description of this procedure, together with the associated
figures demonstrating the consistency of the Block~N implementation,
is provided in Appendix~\ref{app:blockN_validation}.

\vspace{0.5em}
\noindent\textbf{Infrared safety and stability}\\
A well-known feature of QCD perturbative calculations at fixed order is 
singularities in the soft and collinear limits for real emission
matrix elements. 
\\
In our implementation, this cancellation is handled at the level of the
matrix elements provided by \textsc{POWHEG-BOX-V2}, which we modified to supply
finite, infrared-safe Born+virtual and real-emission contributions at each
phase-space point.
Nevertheless, embedding these ingredients into a fully differential,
multidimensional MEM integration requires additional care from a
numerical point of view.
\\
\\
To ensure stable and efficient numerical convergence of the
MEM integral, we impose a minimal transverse-momentum threshold on the
additional radiated parton within the integration.
This requirement does not constitute a subtraction of infrared
singularities nor a modification of the underlying NLO prediction, but
serves as a technical regulator to avoid regions of phase space
that lead to inefficient sampling.

\vspace{0.5em}
\noindent\textbf{Collinear treatment module}\\
In addition to the transverse-momentum cut-off, a dedicated collinear
treatment module is implemented to regulate configurations in which the
radiated parton becomes nearly parallel to an incoming gluon or one of
the Higgs decay products. In this collinear limit, the real-radiation
four-momentum, together with that of the collinear particle, is locally
redefined to reflect how the detector would reconstruct only the decay
product. 
This procedure also ensures smooth numerical behaviour of the MEM 
integrand across the singular regions.

\section{Validation and Discrimination Power} 

\begin{figure}[h]
  \centering
  \includegraphics[width=0.48\textwidth]{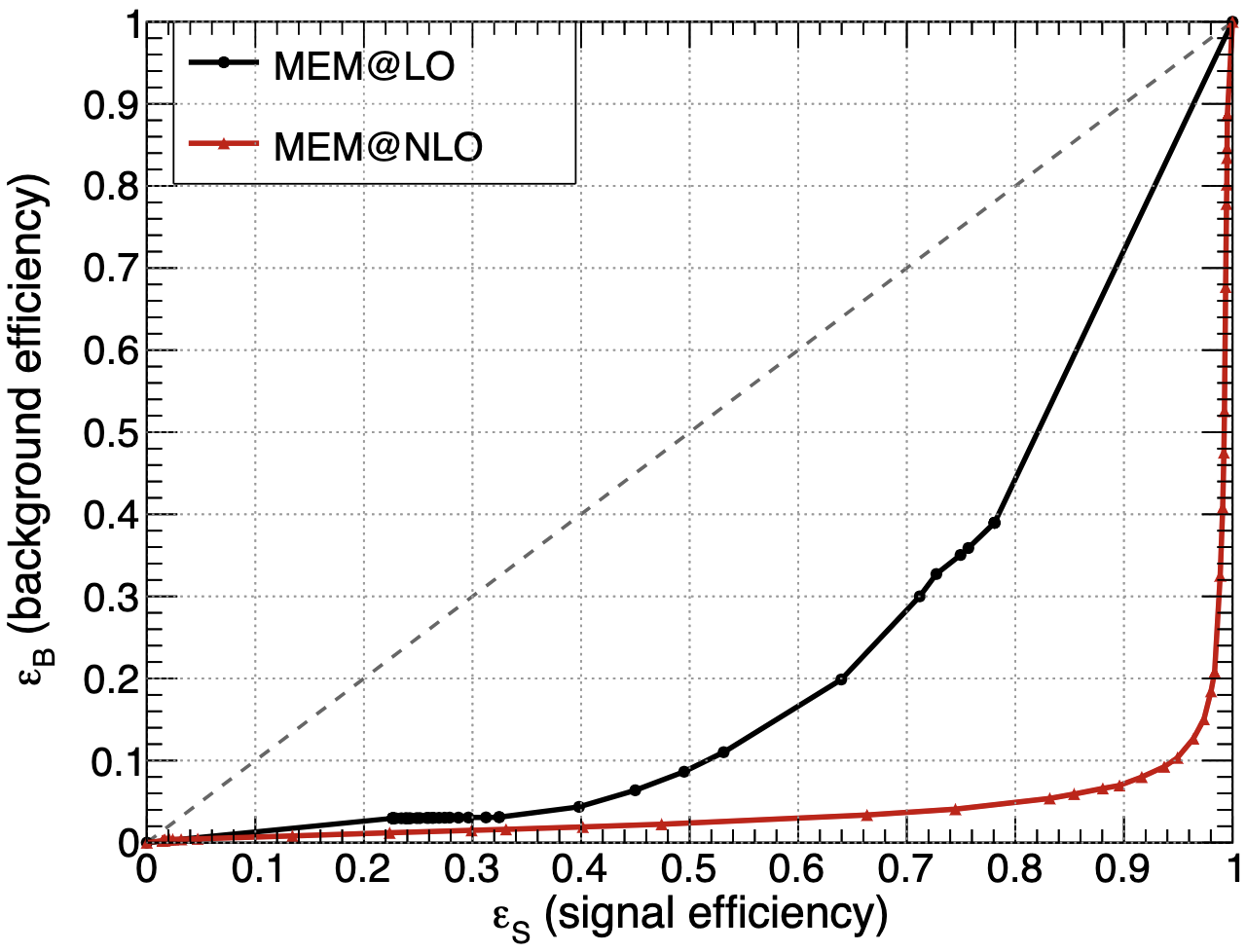}
  \caption{ROC curve for events generated at NLO, between the $ggF$ di-Higgs signal and ttH background. 
  The black curve corresponds to MEM@LO, and the red curve to MEM@NLO.
  Given our conventions, the closer the ROC curve lies to
  the bottom-right corner of the plot, the more powerful the method.}
  \label{fig:ROC}
\end{figure}

The separation power of the MEM@NLO is most directly illustrated with
rejection curves (ROC curves), built from the
event-by-event likelihood ratio between the signal and a chosen background
hypothesis. The efficiency of a given cut on this discriminant is
evaluated using simulated samples of signal (horizontal axis) and background (vertical axis) events.
For this analysis, the likelihoods are computed from Eq.~\eqref{eq:likelihood} using the MEM.
\\
\\
The comparison in Fig.~\ref{fig:ROC} highlights the necessity of
extending the MEM to NLO accuracy.  
The MEM@LO has its sensitivity degrading significantly on realistic NLO
samples as it does not account for the extra radiation.
In contrast, the MEM@NLO achieves and restores an improved discrimination power on NLO
events, thanks to its more realistic description of the underlying
physics.
\\
\\
Beyond the ROC performances, additional cross-checks were performed to
validate the numerical accuracy of the MEM@NLO implementation and its many interfaces.
The LO and real-emission matrix elements obtained from
\textsc{POWHEG-BOX-V2} were compared with those from
\textsc{MadGraph5\_aMC@NLO}~\cite{MadGraph_Article} (another broadly used Monte Carlo generator), 
showing excellent agreement 
across representative phase-space points.

\section{Extraction of the Higgs Self-Coupling using likelihood scans}

Beyond discrimination between the different processes~$p$, 
the ultimate goal of the MEM@NLO is to extract the
Higgs trilinear coupling $\kappa_{\lambda}$. 
To achieve this, one can define three likelihood functions:

\begin{itemize}
  \item The \emph{kinematic likelihood},
      $\mathcal{L}_{\text{kin}}(\kappa_{\lambda})$, which compresses the
      full event–level kinematic information into a single likelihood for
      the entire dataset:

      (i) Process likelihood. 
      For each selected event $i$ and each process $p$ (signal and, 
      separately, each of the background processes) 
      the process-level likelihood is evaluated using
      Eq.~\eqref{eq:likelihood}:
      \[
        \mathcal{L}^{p}_{\text{process}}
        (\kappa_{\lambda}\mid\mathbf{x}^{\,i}) .
      \]

      (ii) Event likelihood.  
      The previous contributions are combined into the per–event likelihood
      through the weighted sum for a given event $i$:
      \begin{equation}
        \mathcal{L}_{\text{event}}(\kappa_{\lambda}\mid\mathbf{x}^{\,i})
        = \sum_{p=1}^{n_{P}}
          f_{p}(\kappa_{\lambda})\,
          \mathcal{L}^{p}_{\text{process}}
          (\kappa_{\lambda}\mid\mathbf{x}^{\,i}),
        \label{eq:event_like}
      \end{equation}
      where $f_{p}$ denotes the fraction of events in the sample that
      originates from process $p$ (with $\sum_{p} f_{p}=1$), fixed here
      using the theoretically expected signal and background compositions.

      \newpage
      (iii) Kinematic (or sample) likelihood.  
      The kinematic likelihood for the full dataset is then obtained as
      the product of the event likelihoods:
      \begin{equation}
        \mathcal{L}_{\text{kin}}(\kappa_{\lambda})
        = \prod_{i=1}^{N}
          \mathcal{L}_{\text{event}}(\kappa_{\lambda}\mid\mathbf{x}^{\,i}),
        \label{eq:kinematic_like}
      \end{equation}
      fully encoding the event–level kinematics of the selected sample.
  \item The \emph{yield likelihood} $\mathcal{L}_{\text{yield}}(\kappa_{\lambda})$,
        exploits the strong dependence of the $gg\!\to\!HH$ production
        rate on $\kappa_{\lambda}$ which is known to vary quadratically due to the interference of
        the triangle and box diagrams. 
        \\
        The yield likelihood follows a Poisson distribution, 
        with parameter $\mu(\kappa_{\lambda}) = N_{\rm sig}(\kappa_{\lambda}) + N_{\rm bkg}$, being the expected event yield, 
        and $N_{\text{obs}}$ being the observed number of events:
        \begin{equation}
          \mathcal{L}_{\text{yield}}(\kappa_{\lambda}) = \frac{\mu(\kappa_{\lambda})^{N_{\text{obs}}} e^{-\mu(\kappa_{\lambda})}}{N_{\text{obs}}!}.
          \label{eq:yield_like}
        \end{equation}
  \item The \emph{extended likelihood} $\mathcal{L}_{\text{ext}}$, defined as
      the product of the kinematic and yield components,
      \begin{equation}
        \mathcal{L}_{\text{ext}}(\kappa_{\lambda})
        = \mathcal{L}_{\text{kin}}(\kappa_{\lambda})\,
          \times\mathcal{L}_{\text{yield}}(\kappa_{\lambda}),
        \label{eq:extended_likelihood}
      \end{equation}
      exploiting simultaneously the full event-level kinematic
      information and the overall dependence of the event yield on
      $\kappa_{\lambda}$.
\end{itemize}

\vspace{2em}
\noindent\textbf{Negative Log-Likelihood}\\
For the statistical extraction of $\kappa_{\lambda}$, it is convenient to
work with the negative log-likelihood (NLL), which transforms the product over
events in Eq.~\eqref{eq:extended_likelihood} into a numerically stable sum:
\begin{equation}
\begin{aligned}
  -\ln \mathcal{L}_{\text{ext}}(\kappa_{\lambda})
  &= -\sum_{i=1}^{N_{\rm obs}}
     \ln \mathcal{L}_{\text{event}}(\kappa_{\lambda}\mid\mathbf{x}^{\,i})
\\[4pt]
  &\quad +\, \mu(\kappa_{\lambda})
     - N_{\rm obs}\,
       \ln\!\big[\mu(\kappa_{\lambda})\big]
     + C ,
\end{aligned}
\label{eq:nll_ext}
\end{equation}

\noindent
where $C$ is a constant term that does not affect the minimisation of the NLL.
\\
\\
The NLL provides a test statistic whose minimum identifies the best-fit value
$\hat{\kappa}_{\lambda}$, while its curvature around the minimum yields
the corresponding statistical uncertainty. 
\\
This formulation is
equivalent to the standard likelihood-ratio approach used in high-energy
physics and is well suited for profiling $\kappa_{\lambda}$
in large ensembles of pseudo-experiments.

\newpage
\noindent\textbf{Likelihood scans and associated histograms}\\

\begin{figure}[htbp!]
  \centering
  \includegraphics[width=0.5\textwidth]{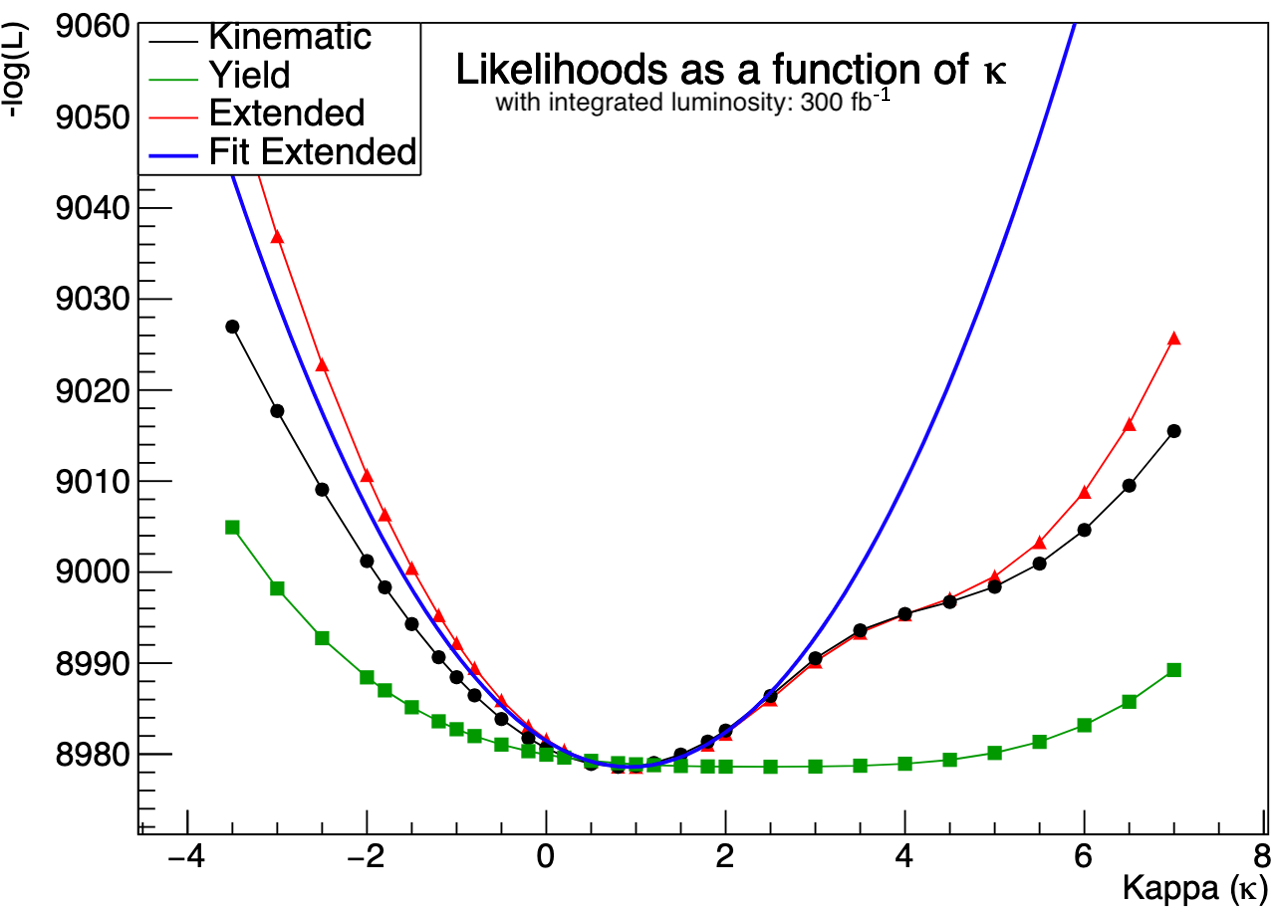}
  \caption{
    Likelihood scan for one pseudo–experiment consisting of a combination of NLO 
    signal events ($\kappa_{\lambda}=1$) and all main backgrounds, using MEM@NLO.
    The graph shows \(-\log\mathcal{L}\) as a function of the hypothesis~\(\kappa_{\lambda}\).
    \textbf{Black:} kinematic component (\(\mathcal{L}_{\text{kin}}\)) with offset;  
    \textbf{\textcolor{green}{Green}:} yield term (\(\mathcal{L}_{\text{yield}}\)) with offset;  
    \textbf{\textcolor{red}{Red}:} extended likelihood (\(\mathcal{L}_{\text{ext}}\));  
    \textbf{\textcolor{blue}{Blue}:} quadratic fit to \(-\log\mathcal{L}_{\text{ext}}\) around its minimum.
    }
  \label{fig:LikelihoodScan}
\end{figure}

\begin{figure}[h]
  \centering
  \includegraphics[width=0.58\textwidth]{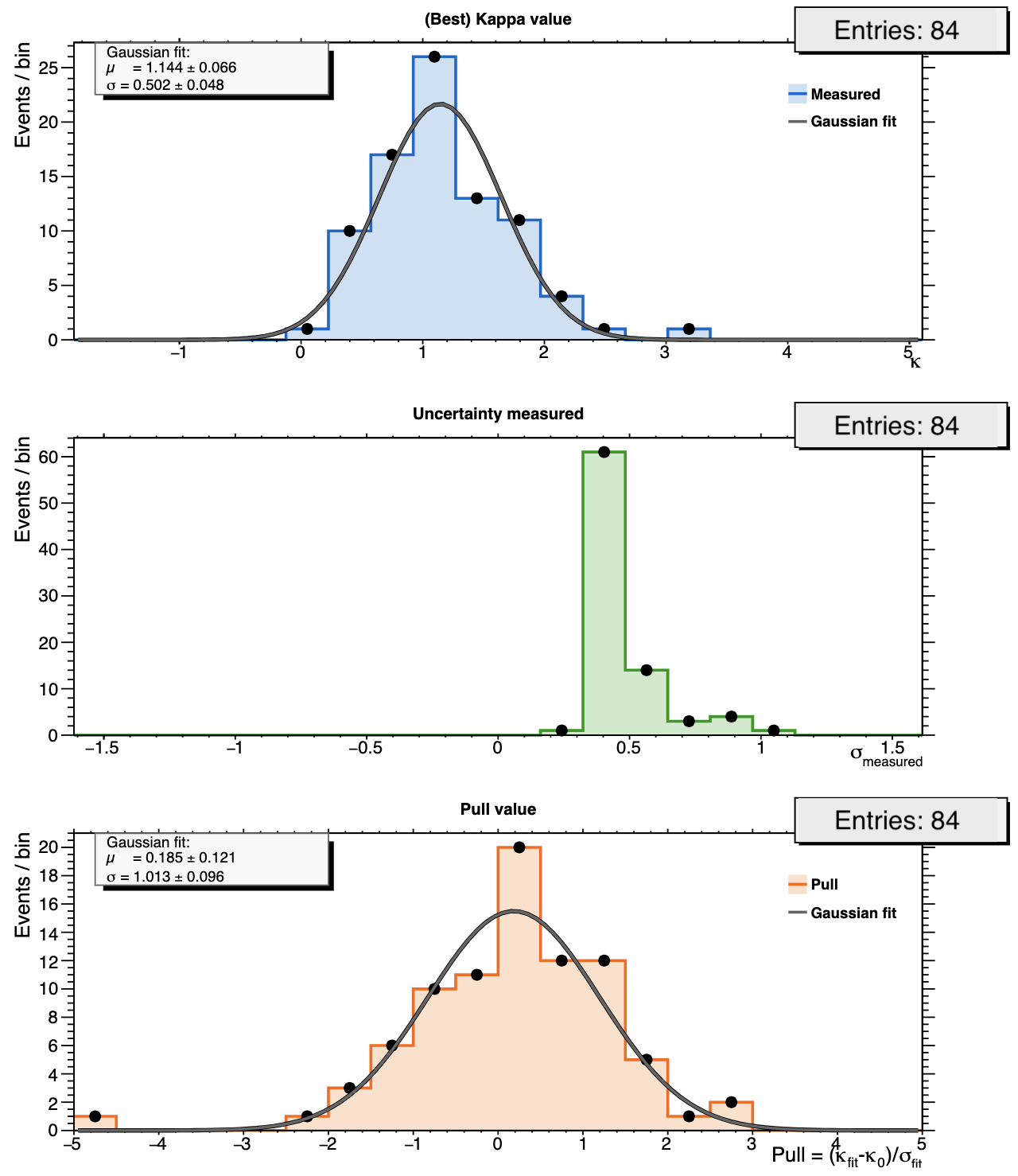}
  \caption{Histograms of likelihood scan results of pseudo-experiments including signal and all main backgrounds. 
  Top: distribution of the best-fit coupling parameter $\hat{\kappa_{\lambda}}$, with Gaussian fit overlaid. 
  Middle: distribution of the estimated uncertainty $\sigma_{\text{measured}}$.
  Bottom: distribution of the pull values $\omega=(\hat{\kappa_{\lambda}} - \kappa_{\lambda}^{\text{true}})/\sigma_{\text{measured}}$.}
  \label{fig:LikelihoodPerformance}
\end{figure}

\noindent
To quantify the performance of the MEM@NLO, an ensemble of 84 independent
pseudo-experiments was generated at NLO accuracy, each corresponding to an
integrated luminosity of 300~fb$^{-1}$ and assuming the Standard Model
value $\kappa_{\lambda}=1$.
\\
\\
Each pseudo-experiment is constructed by fluctuating the expected signal
and background yields according to Poisson statistics and by drawing
events from the corresponding NLO Monte Carlo samples.
\\
For each pseudo-experiment, the extended likelihood
$\mathcal{L}_{\text{ext}}(\kappa_\lambda)$ is evaluated for a large number of values of 
$\kappa_\lambda$. 
The result for one of the pseudo-experiments is shown in Fig.~\ref{fig:LikelihoodScan}.
\\
The position of the minimum of the profile
$-\log\mathcal{L}_{\text{ext}}(\kappa_\lambda)$ defines the best-fit value
$\hat{\kappa}_\lambda$, while the local curvature around the minimum
provides an estimate of the statistical uncertainty associated with that
pseudo-experiment.
\\
\\
Repeating this procedure over the full set of pseudo-experiments allows
one to build distributions of the fitted parameter
$\hat{\kappa}_\lambda$ and the estimation of the corresponding uncertainty.
\\
These distributions are used to assess any bias and the expected
statistical uncertainty of the measurement of $\kappa_{\lambda}$ using
our implementation of the MEM@NLO. The distributions obtained from
Monte Carlo pseudo-experiments including signal and all main
backgrounds are shown in Fig.~\ref{fig:LikelihoodPerformance}.

\newpage
\vspace*{0.2cm}

\noindent
In addition, we show the distribution of the pull, defined as 
$\omega = (\hat{\kappa}_\lambda-\kappa_{\lambda}^{\text{true}})
/\sigma_{\text{measured}}$. 
The pull is used to validate the reliability of the estimated statistical uncertainty.
\\
The observed pull distribution is centered around zero and exhibits 
a width close to unity, indicating that any bias resulting from the implementation of the MEM and 
the likelihood-based procedure is reasonably small, and that the reported
uncertainties are reliable estimates.

\clearpage

\section{Results}

\begin{figure*}[h]
  \centering
  \includegraphics[width=0.82\textwidth]{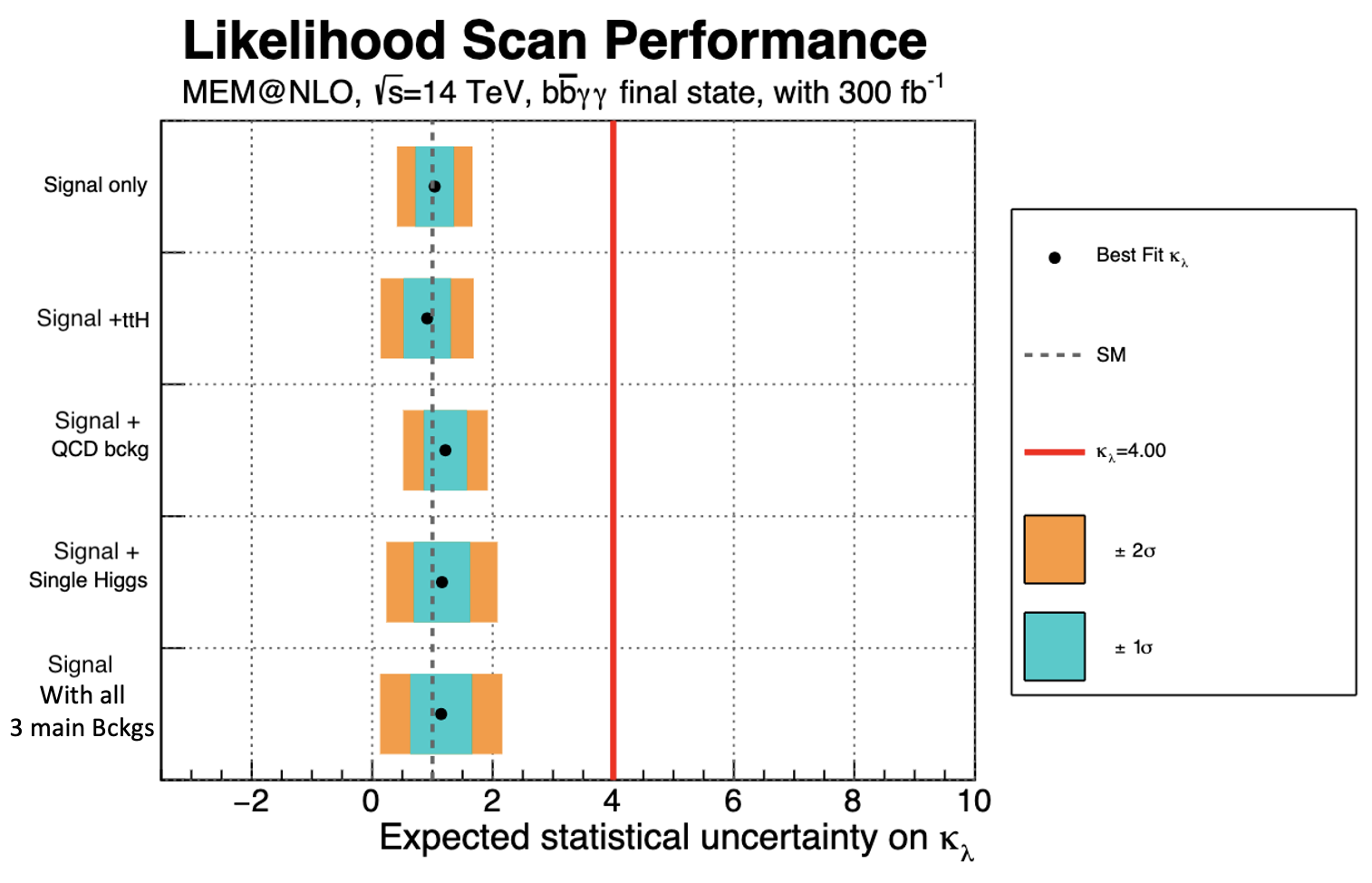}
  \caption{Summary of likelihood scan performance for MEM@NLO on NLO
  Monte Carlo pseudo-experiments generated with $\kappa_\lambda=1$ in the $b\bar{b}\gamma\gamma$ final state, with 300~$fb^{-1}$. 
  The black dot marks the mean best-fit value of $\kappa_\lambda$ for each set of pseudo-experiments, 
  while the bands show the statistical $\pm 1\sigma$ (turquoise) and $\pm 2\sigma$ (orange)
  ranges. The dashed grey line is the SM prediction ($\kappa_\lambda=1$), and the red
  vertical line the location of the mirror solution near $\kappa_\lambda=4$.}
  \label{fig:summary}
\end{figure*}

To assess the impact of the different main backgrounds on the expected 
statistical uncertainty of a measurement of $\kappa_\lambda$ using the MEM@NLO, 
we produce and analyse multiple samples with different background compositions: 
a signal-only sample, samples including each background process
individually, and a sample combining all major background sources
simultaneously.  
The results are summarised in
Fig.~\ref{fig:summary} and in Table~\ref{tab:NLO_likelihood_summary}.  
Across all configurations, the injected Standard Model value
$\kappa_\lambda = 1$ is recovered, with acceptable bias (corrections which can be included in a future analysis of collider data). 
For the analysis of 84 pseudo-experiments, including \textbf{all main} backgrounds, which offers the most realistic environment, 
the likelihood fit yields
$\kappa_\lambda = 1.144 \pm 0.502$ where $1.144$ represents the mean over all pseudo-experiments, 
and $0.502$ the expected 1-$\sigma$ uncertainty 
at 300~fb$^{-1}$. 
\\
This result demonstrates that the MEM@NLO retains strong
sensitivity, even in a challenging background-dominated environment.
Another important feature of the MEM@NLO is its ability to remove the
mirror solution near $\kappa_\lambda = 4$. This secondary solution
appears in yield-based analyses because the inclusive $gg \to HH$ cross
section depends on the interference between the triangle and box
diagrams, allowing this second value of $\kappa_\lambda$ to give the
same total rate as the SM. By exploiting the full
event-level kinematic information, MEM@NLO breaks this degeneracy and
selects only the physical solution.

\onecolumngrid

\begin{table*}[h]
  \caption{
    Summary of likelihood-scan performances for NLO generated events analyzed with MEM@NLO.
    Each background process listed is included on top of the signal (signal+background).
    This table corresponds to Fig.~\ref{fig:summary}.
    Uncertainties in this table reflect the statistical power of the pseudo-experiment sets.
  }
  \centering
  \renewcommand{\arraystretch}{1.15}
  \makebox[\textwidth][c]{%
  \begin{tabular}{|l||c|c||c||c|c|}
    \hline
    \multirow{2}{*}{\textbf{Process}} &
    \multicolumn{2}{c||}{\textbf{Best fit}} &
    \multirow{2}{*}{\parbox{2.2cm}{\centering \textbf{Measured} \\ \textbf{uncertainty}}} &
    \multicolumn{2}{c|}{\textbf{Pull distribution}} \\
    \cline{2-3 } \cline{5-6}
    & $\kappa_{\text{mean}}$ & $\sigma_{\kappa}$ &
    & $\omega_{\text{mean}}$ & $\sigma_{\text{pull}}$ \\
    \hline\hline
    Signal only
      & $1.037 \pm 0.012$ & $0.309 \pm 0.008$
      & $0.302 \pm 0.030$
      & $0.123 \pm 0.039$ & $1.020 \pm 0.026$ \\
    \hline
    $t\bar tH$
      & $0.913 \pm 0.042$ & $0.441 \pm 0.026$
      & $0.381 \pm 0.087$
      & $-0.391 \pm 0.113$ & $1.422 \pm 0.092$ \\
    \hline
    QCD $\, (b\bar b\gamma\gamma)$
      & $1.215 \pm 0.040$ & $0.354 \pm 0.027$
      & $0.345 \pm 0.030$
      & $0.555 \pm 0.126$ & $1.023 \pm 0.103$ \\
    \hline
    Single Higgs
      & $1.158 \pm 0.039$ & $0.478 \pm 0.030$
      & $0.418 \pm 0.088$
      & $0.296 \pm 0.085$ & $1.044 \pm 0.058$ \\
    \hline
    \textbf{With all backgrounds}
      & $\mathbf{1.144 \pm 0.066}$ & $\mathbf{0.502 \pm 0.048}$
      & $\mathbf{0.485 \pm 0.142}$
      & $\mathbf{-0.185 \pm 0.121}$ & $\mathbf{1.013 \pm 0.096}$ \\
    \hline
  \end{tabular}}%
  \label{tab:NLO_likelihood_summary}
\end{table*}

\clearpage
\twocolumngrid
\newpage

\section{Computational cost and scalability}
\label{sec:computational_cost}

\begin{figure}[h]
  \centering
  \includegraphics[height=6.5cm]{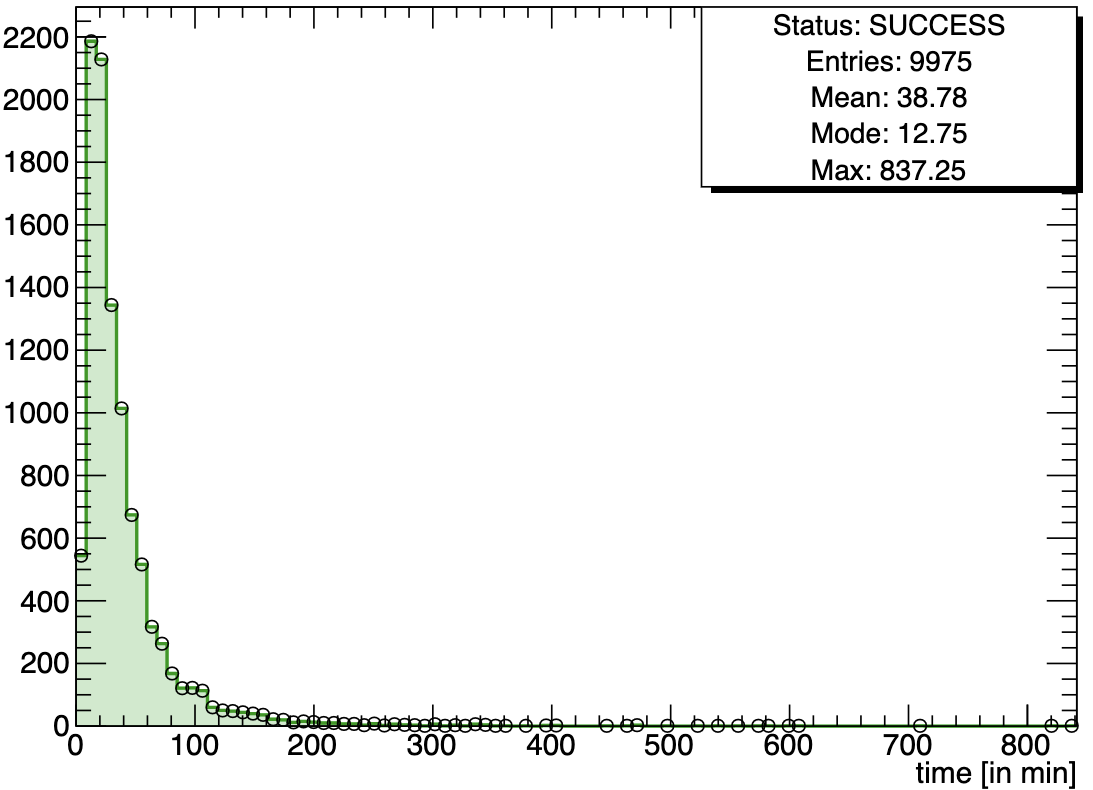}
  \caption{Wall-clock Time (in minutes) for the calculation of per-event 
          $\mathcal{L}^{p}_{\text{process}}$ for an NLO signal sample evaluated 
           for the signal process.
           The mean, mode and maximum values of the distributions are indicated.}
  \label{fig:Time_MEM}
\end{figure}

As with all applications of the Matrix Element Method, the numerical
evaluation of the likelihood involves high-dimensional integrations
over the phase space, leading to a substantial computational
cost.
This feature is intrinsic to any MEM approach, where precision is
obtained by integrating over all kinematic configurations compatible
with a given reconstructed event (\textit{cf}. Eq.~\eqref{eq:likelihood}).
\\
\\
The extension of the MEM to next-to-leading order further increases this
complexity with the addition of three unresolved
radiation degrees of freedom for the real contributions.
\\
The introduction of \textsc{Block~N} allows this integration to be
performed in a controlled and numerically stable way, but it does not
remove the fundamental scaling of the problem with dimensionality.
\\
The dimensionality of the phase-space integrations
entering the MEM@NLO evaluation for each signal and background process
considered in this study is summarised in Appendix~\ref{app:dof_table}.
\\
\\
In the present study, the performance of the \textsc{MoMEMta} framework
was optimised as far as possible within its existing design, including
careful choices of integration strategies and configuration of the numerical 
integration algorithms implemented in \textsc{CUBA}~\cite{CUBA_paper}, the software for integration 
that is interfaced to \textsc{MoMEMta} by default. 
Nevertheless, the computational cost remains significant.
\\
The distribution of the wall-clock time required to evaluate the MEM@NLO likelihood for the 
signal process\footnote{on a single thread of an AMD EPYC\texttrademark~7453 processor, at 2.75 GHz with 28 cores running 56 threads.} is shown in Fig.~\ref{fig:Time_MEM}. On average it is 38.8 minutes.
Yet for a small fraction of events, the wall-clock was found to reach several
hundred minutes, with extreme cases approaching
$838$~minutes.
\\
\\
Looking forward, several promising directions exist to improve the
practical scalability of MEM@NLO.
\\
Recent developments in machine-learning–assisted importance sampling
and adaptive integration techniques offer the possibility to guide the
phase-space exploration toward the most relevant regions, thereby
significantly reducing the number of integrand evaluations required. 
A recent discussion of these approaches can be found \textit{e.g.} in Ref.~\cite{Ubiali} 
as well as the references therein~\cite{ref_1sur6,ref_2sur6,ref_3sur6,ref_5sur6,
ref_6sur6,ref_7sur6,ref_8sur6,ref_9sur6,ref_10sur6}.
\\
\\
In addition, the use of surrogate models to emulate the calculation 
of the likelihood~\cite{Jan_article_1sur3,Jan_article_2sur3,Jan_article_3sur3} will facilitate systematic studies of
theoretical and experimental uncertainties without repeated full
re-evaluations.
\\
While these developments are beyond the scope of the present work, they
provide a natural continuation of the MEM@NLO framework introduced
here, and will be essential for its application in future high-precision analyses of 
LHC and HL-LHC data.

\section{Conclusions and Outlook}
We present the first implementation of the MEM at NLO for the extraction of 
the Higgs boson self-coupling $\lambda_{3H}$ in Higgs boson pair
production in the $b\bar b\gamma\gamma$ channel. 
A central achievement of this work is the development of
\textsc{Block~N}, a new integration block specifically designed to
treat the additional degrees of freedom associated with real emission contributions at NLO.  
Its construction required a substantial extension of the
existing \textsc{MoMEMta} formalism and enabled for the first time a
consistent NLO treatment, resolving the limitations of previous LO-based approaches.
\\
\\
Validation on simulated samples demonstrates significant improvements
in discrimination power and parameter extraction. Ensemble tests
confirmed the robustness of the method and its ability to break the
mirror solution near $\kappa_{\lambda}=4$.
The method provides a powerful framework for extracting the
Higgs self-coupling and can be naturally extended to other analyses or
processes.
\\
\\
Future development of this framework could focus on refining the modelling 
of detector transfer functions and exploring the application of MEM at NLO 
to a broader range of final states such as 
$b\bar b\tau^{+}\tau^{-}$ or $b\bar b b\bar b$.
\\
\\
As with all applications of the Matrix Element Method, the computational
cost associated with multidimensional integrations is known to be substantial.
Recent developments in machine-learning-based importance sampling and
adaptive integration could also offer an interesting direction for future studies, 
as they could considerably reduce evaluation time while preserving the
accuracy of the method. 
\\
A further natural extension concerns the
systematic treatment of uncertainties.
Surrogate model approaches capable of emulating the behaviour of the
full likelihood are expected to provide an efficient framework and enhance the use of MEM at NLO in future analyses.

\begin{acknowledgments}
We gratefully acknowledge the support of our colleagues at the
\textbf{IN2P3 Computing Centre} (CC-IN2P3) in Lyon (Villeurbanne) for the
reliable and efficient operation of their computing farms.
Without this facility, the present study would not have been possible.
\end{acknowledgments}


\newpage

\appendix
\section{Validation of Block~N}
\label{app:blockN_validation}

The introduction of \textsc{Block~N} constitutes a central extension of
the \textsc{MoMEMta} formalism at NLO, enabling for the first time the treatment
of real-emission contributions within the Matrix Element Method.
Because this construction is not covered by existing implementations, a dedicated
validation procedure was required.

\vspace{0.5em}
\noindent\textbf{Validation strategy}\\
The validation is based on a direct and physically motivated comparison
between the behaviour of the MEM integrand constructed with
\textsc{Block~N} and the kinematic properties of real-emission events
generated at NLO.
\\
\\
The longitudinal momentum of the unresolved real-emission parton,
$p_z^{\mathrm{rad}}$, is a natural probe of the Block~N construction, as
it is the only remaining integration variable associated with the extra
parton after enforcing four-momentum conservation.
\\
For a large sample of NLO signal events generated with
\textsc{POWHEG-BOX-V2}, the distribution of $p_z^{\mathrm{rad}}$ is
compared to the behaviour of the MEM integrand evaluated at the same
phase-space points using \textsc{Block~N}.
This comparison allows one to assess whether the MEM integrand correctly
reproduces the physical distribution of the unresolved radiation.

\begin{figure*}[t]
  \centering
  \begin{tabular}{cc}
    \parbox[c]{0.49\textwidth}{
      \centering
      \includegraphics[width=\linewidth]{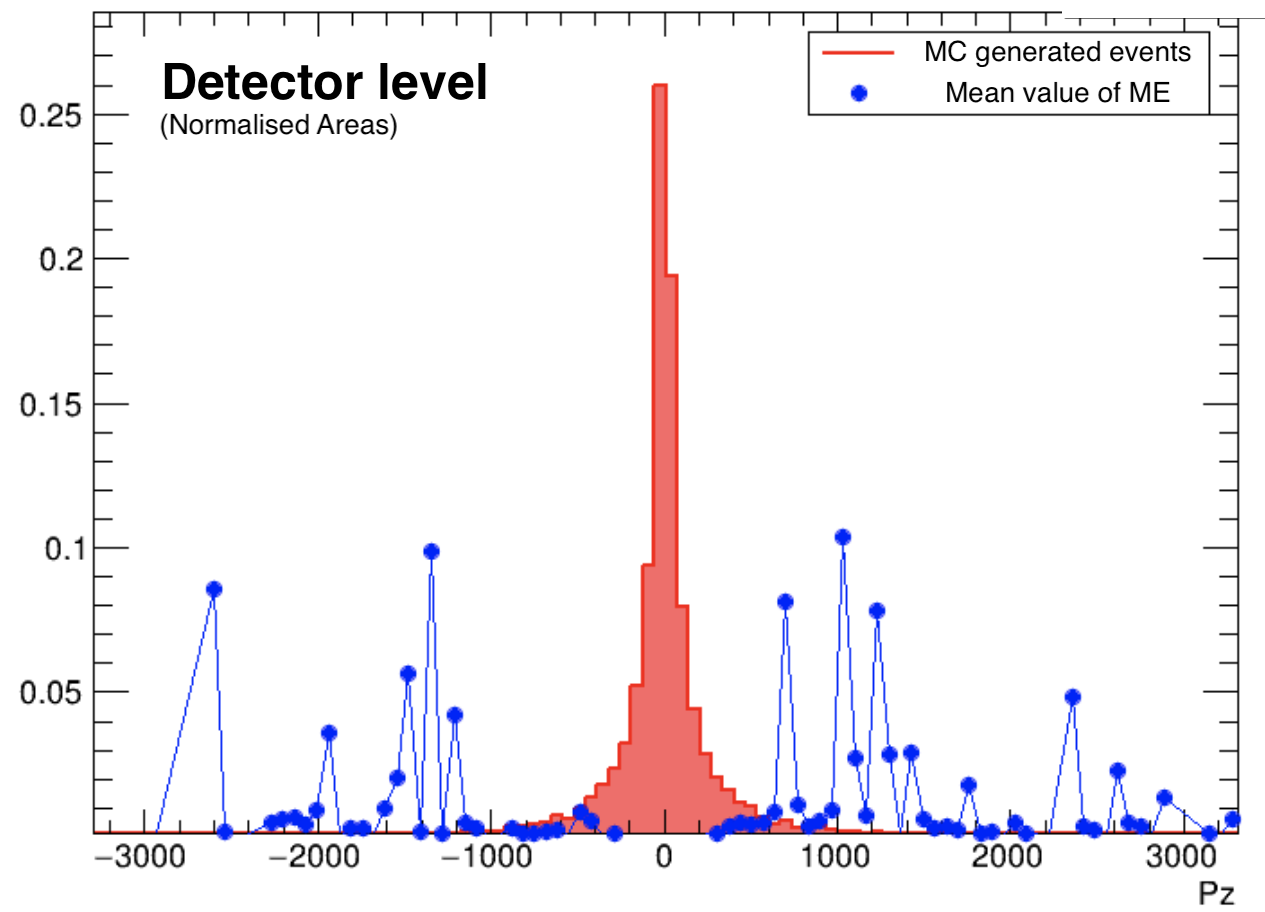}\\
      \small a) Real matrix element only
    } &
    \parbox[c]{0.49\textwidth}{
      \centering
      \includegraphics[width=\linewidth]{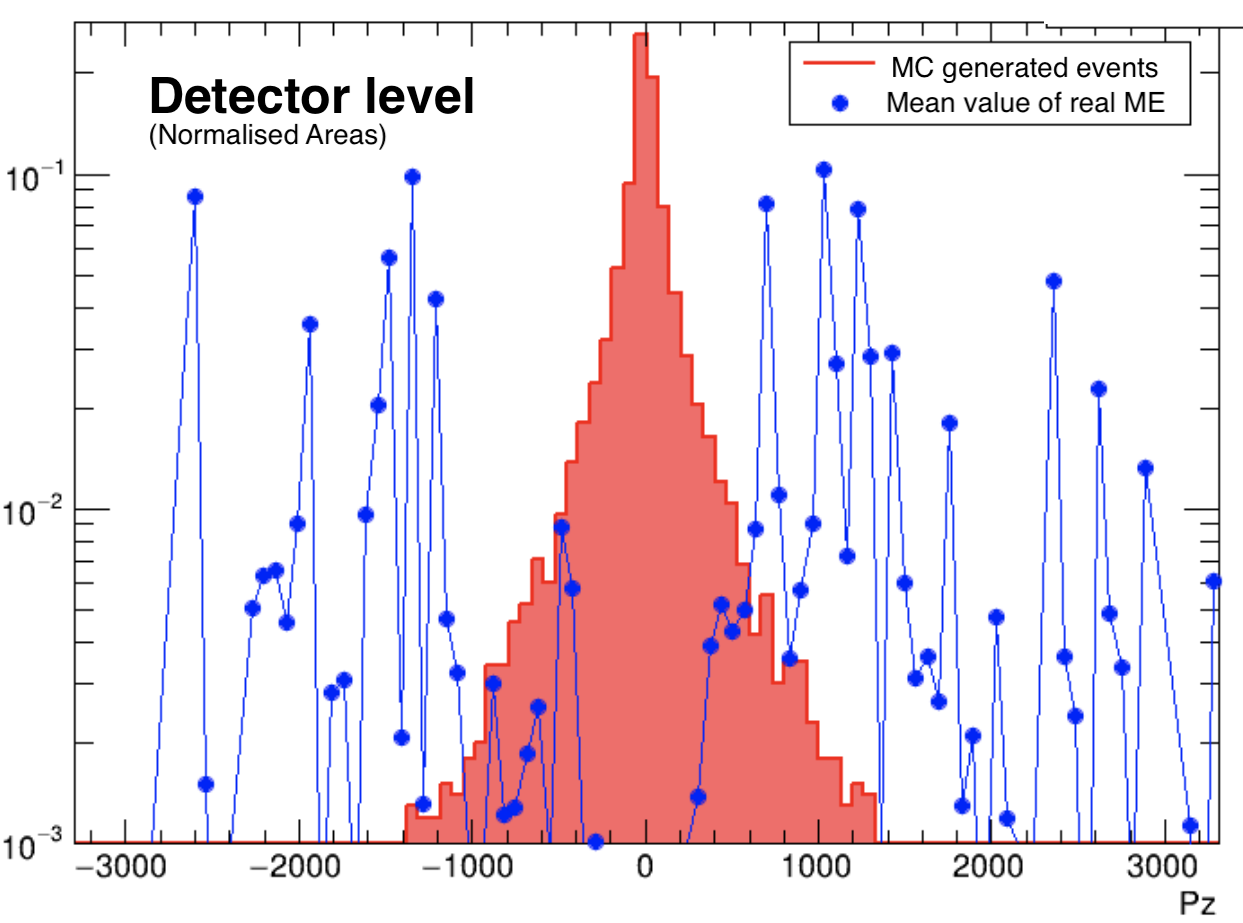}\\
      \small b) Real matrix element only (vertical axis on logarithmic scale)
    } \\[1ex]
    \parbox[c]{0.49\textwidth}{
      \centering
      \includegraphics[width=\linewidth]{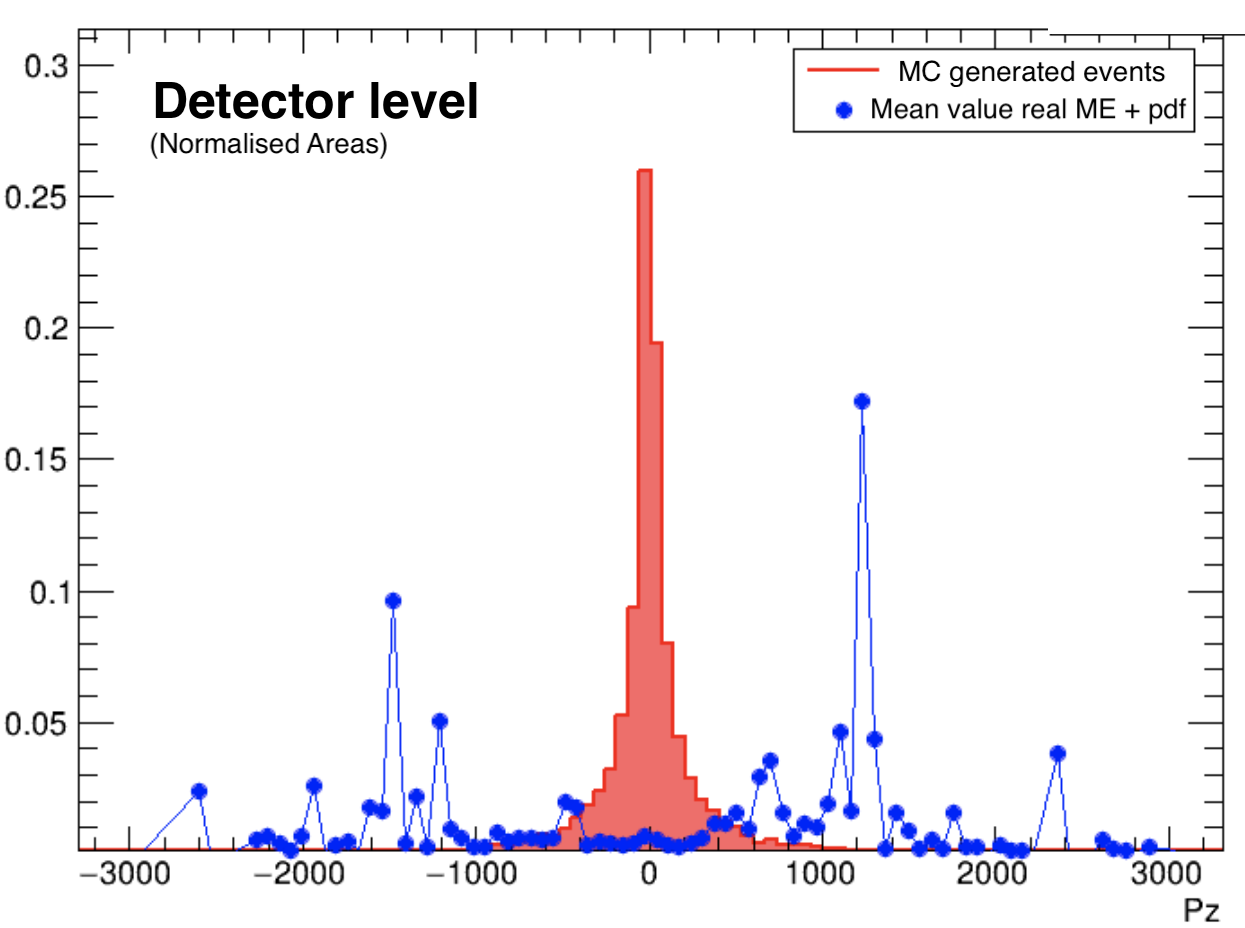}\\
      \small c) Real matrix element weighted by PDFs
    } &
    \parbox[c]{0.49\textwidth}{
      \centering
      \includegraphics[width=\linewidth]{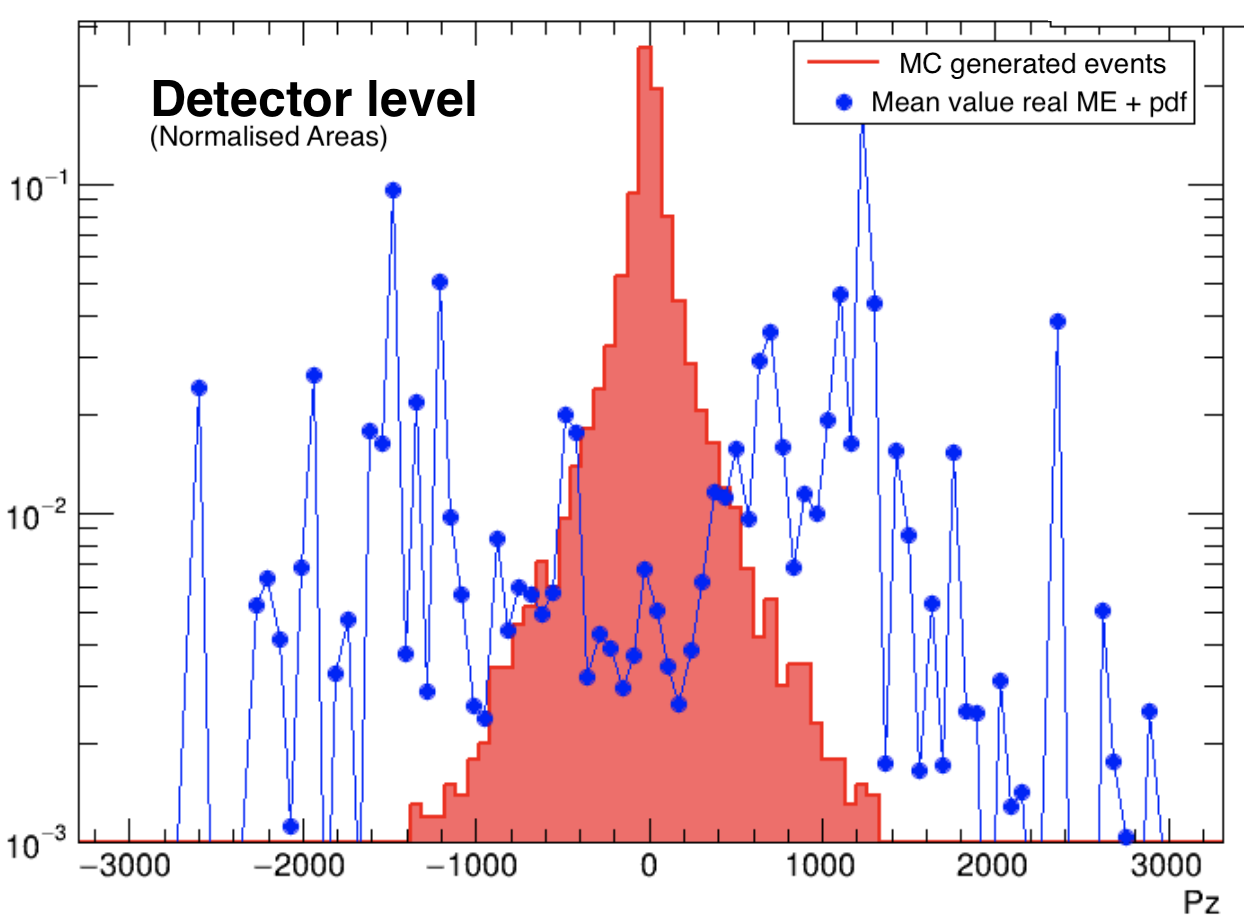}\\
      \small d) Real matrix element weighted by PDFs (vertical axis on logarithmic scale)
    } \\[1ex]
    \parbox[c]{0.49\textwidth}{
      \centering
      \includegraphics[width=\linewidth]{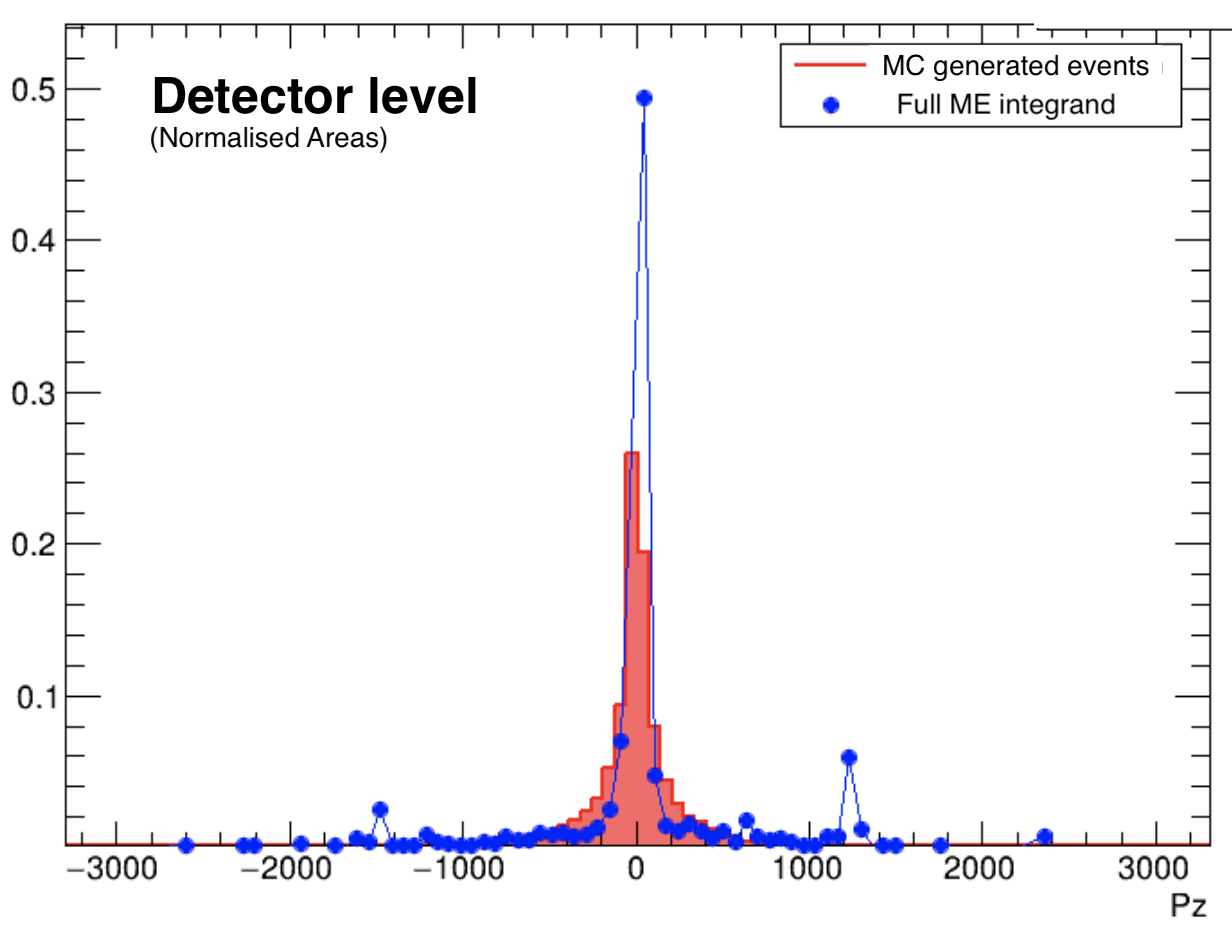}\\
      \small e) Full MEM integrand
    } &
    \parbox[c]{0.49\textwidth}{
      \centering
      \includegraphics[width=\linewidth]{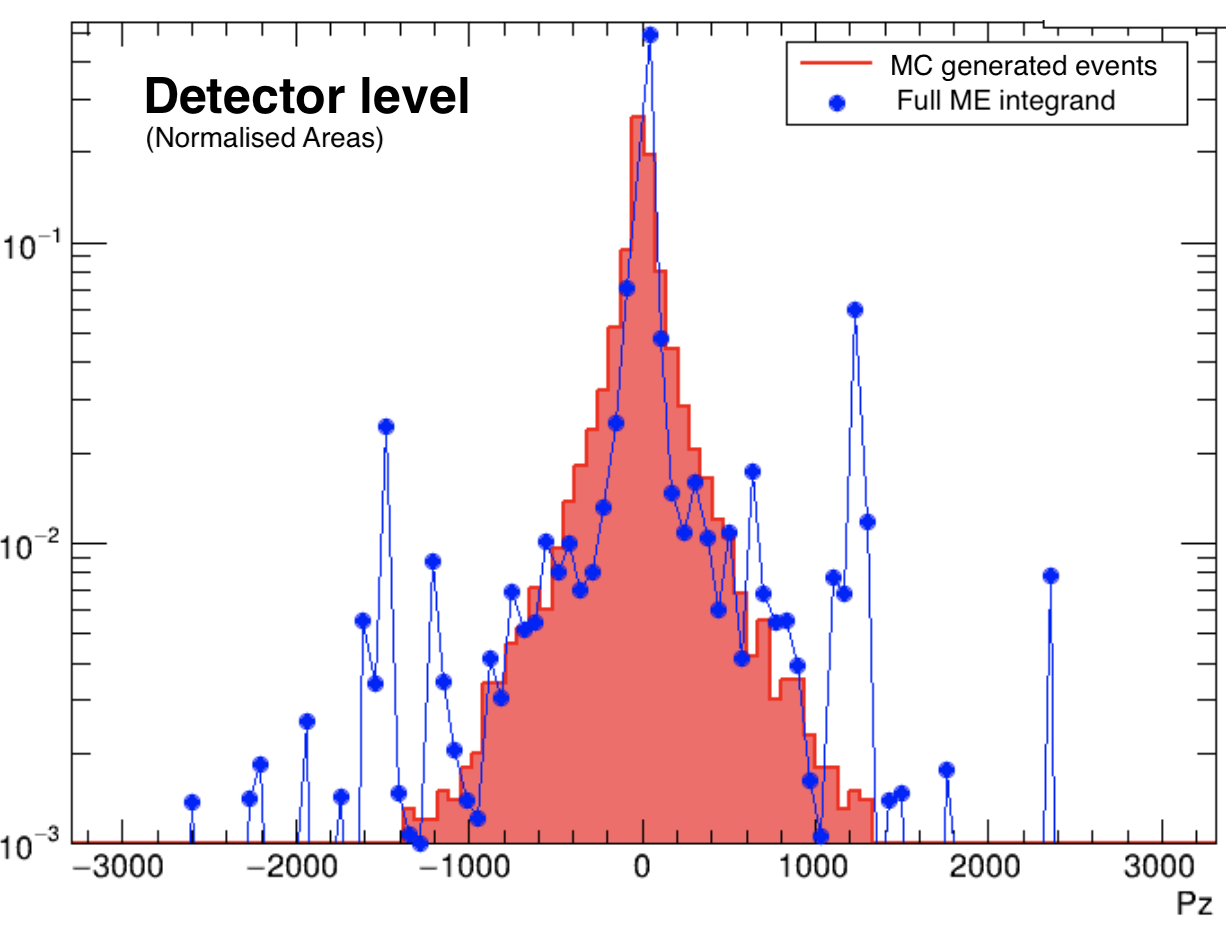}\\
      \small f) Full MEM integrand (vertical axis on logarithmic scale)
    }
  \end{tabular}
  \caption{
    Validation of the \textsc{Block~N} construction using the longitudinal
    momentum of the unresolved real-emission parton $p_z^{\mathrm{rad}}$.
    Panels show the distribution of $p_z^{\mathrm{rad}}$ obtained from
    NLO Monte Carlo events (red) compared to the mean value of the MEM
    integrand evaluated with \textsc{Block~N} (blue), with all distributions
    normalised to unity.
    Results are shown for increasing levels of complexity:
    (a,b) real matrix element only,
    (c,d) real matrix element weighted by parton distribution functions,
    and (e,f) full MEM integrand including the Block~N Jacobian.
    Right panels use a logarithmic vertical scale for the vertical axis.
  }
  \label{fig:Step_Validation_BlockN}
\end{figure*}

\vspace{0.5em}
\noindent\textbf{Progressive inclusion of MEM ingredients}\\
To isolate the impact of each ingredient entering the MEM@NLO
construction, the validation is performed in successive steps of
increasing complexity, shown in Fig.~\ref{fig:Step_Validation_BlockN}:
\begin{itemize}
  \item Real matrix element only
  [Fig.~\ref{fig:Step_Validation_BlockN}(a,b)]:
  the squared real-emission matrix element is evaluated as a function of
  $p_z^{\mathrm{rad}}$, without parton distribution functions or
  Jacobian factors.
  \item Real matrix element with PDFs
  [Fig.~\ref{fig:Step_Validation_BlockN}(c,d)]:
  the same quantity is weighted by the initial-state parton distribution
  functions, introducing the dependence on the Bjorken-$x$ momentum
  fractions.
  \item Full MEM integrand
  [Fig.~\ref{fig:Step_Validation_BlockN}(e,f)]:
  the complete integrand is constructed by combining the real matrix
  element, the parton distribution functions, and the full Jacobian
  associated with the \textsc{Block~N} transformation and the
  phase-space measure.
\end{itemize}

\noindent
At each stage, the mean of the quantity under study (real matrix element, ..., full MEM integrand) 
is shown in bins of $p_z^{\mathrm{rad}}$, and compared to the distribution of 
$p_z^{\mathrm{rad}}$ in red from the NLO Monte Carlo events.
All distributions are normalised to unity to facilitate a direct
comparison of their shapes.

\vspace{0.5em}
\noindent\textbf{Results and interpretation}\\
The comparison shows that neither the real matrix element alone nor its
PDF-weighted form is sufficient to reproduce the physical
$p_z^{\mathrm{rad}}$ distribution.
\\
In both cases, sizable distortions appear, indicating that large regions
of the integration domain are improperly weighted.
\\
Only after the inclusion of the full Jacobian associated with this configuration 
(partially derived in section~\ref{sec:inter_mom} for the \textsc{Block~N} contribution only) does the MEM integrand accurately follow the Monte Carlo
distribution over the full range of $p_z^{\mathrm{rad}}$.
\\
\\
This validation confirms that \textsc{Block~N} provides a physically coherent treatment 
of real-emission phase space within the
MEM@NLO framework.
It ensures that unresolved radiation is integrated consistently,
without biasing the likelihood or distorting its kinematic dependence,
thereby enabling reliable NLO-accurate applications of the Matrix Element Method.

\clearpage
\section{Degrees of Freedom and Integration Dimensionality}
\label{app:dof_table}

For a given reconstructed final state, the number of integration
variables depends on the underlying hard-scattering process, the decay
topology, and the perturbative order considered.
\\
\\
Table~\ref{tab:integration_variables} summarises the effective number of
integration degrees of freedom used in this analysis for the main signal
and background processes, at both leading order (LO) and next-to-leading
order (NLO).
The specific choice of integration variables is not unique, but the
choices listed here were found to provide stable and efficient
numerical convergence.
\\
Following standard practice in MEM analyses, the resolution on the direction of final-state
objects in momentum space is neglected in the calculation of $\mathcal{L}_{\text{process}}$
(\textit{i.e.} the measured directions are used to evaluate the
parton-level matrix element), which substantially reduces the number of
integration variables.
\\
\\
Integration variables mainly correspond to the energies $E$ of reconstructed final-state objects (photons $\gamma$, $b$ jets),
and the Bjorken-$x$ momentum fractions ($q_1$,$q_2$) for the
initial-state partons.
Intermediate resonance widths may be neglected,
leading to a further reduction of the dimensionality of the numerical
integration (this reduction is indicated by an arrow `$\rightarrow$' in the \emph{Integration dimension} column of Table~\ref{tab:integration_variables}).
\\
\\
At NLO, real-emission contributions introduce additional degrees of
freedom associated with an unresolved parton.
These are denoted by the superscript `rad' and correspond to momentum
components of the extra radiation that are fully integrated over in the
MEM likelihood.
The treatment of these additional degrees of freedom is enabled either by the
\textsc{Block~N} module described in the main text, or by the \textsc{ExtraRadiation\_3DOF} module.
\\
\\
We created the \textsc{ExtraRadiation\_3DOF} module to introduces 
three additional independent integration dimensions which can be parameterized either 
in Cartesian coordinates ($p_x$,$p_y$,$p_z$) or in cylindrical phase space 
($p_T$,$\phi$,$E$) depending on the user needs. This specific module has been designed for processes where the presence of
invisible particles at LO (\textit{e.g.} $t\bar{t}H$) already force the choice of main (and secondary)
blocks and consequently forbids the use of \textsc{Block~N}.

\par\medskip
\onecolumngrid

\begin{center}
\begin{minipage}{0.98\textwidth}

\medskip
\renewcommand{\arraystretch}{1.3}
\setlength{\tabcolsep}{6pt}

\resizebox{\textwidth}{!}{%
\begin{tabular}{|l|c|p{0.45\textwidth}|}
  \hline
  \textbf{Process} &
  \textbf{Integration dimension} &
  \textbf{Representative integration variables} \\
  \hline
  $gg\to HH$ @ LO
    & \textcolor{cyan}{\textbf{2}} $(\rightarrow 0)$
    & $(H\text{ width}),\ \gamma_{1,E}$ \\
  \hline
  $t\bar t H$ @ LO
    & \textcolor{cyan}{\textbf{9}} $(\rightarrow 6)$
    & $(H\text{ width},\ \text{top}_1\text{ width},\ \text{top}_2\text{ width}),\newline
      \text{permutation of }(b_3,b_4),\newline
      \gamma_{1,E},\ b_{3,E},\ b_{4,E},\ q_1,\ q_2$ \\
  \hline
  QCD $b\bar b\gamma\gamma$ @ LO
    & \textcolor{cyan}{\textbf{2}}
    & $\gamma_{1,E},\ \gamma_{2,E}$ \\
  \hline
  Single Higgs @ LO
    & \textcolor{cyan}{\textbf{2}}
    & $(H\text{ width}),\ \gamma_{1,E}$ \\
  \hline
  \rowcolor{green!10}
  $gg\to HH$ @ NLO (real)
    & $5\ (\rightarrow \textcolor{cyan}{\textbf{3}})$
    & $(H_1\text{ width},H_2\text{ width}),\ \gamma_{1,E},\ b_{3,E},\ p_z^{\rm rad}$ \\
  \hline
  $t\bar t H$ @ NLO (real)
    & $12\ (\rightarrow\textcolor{cyan}{\textbf{9}})$
    & LO variables $+\ p_T^{\rm rad},\ \phi^{\rm rad},\ E^{\rm rad}$ \\
  \hline
  QCD $b\bar b\gamma\gamma$ @ NLO (real)
    & \textcolor{cyan}{\textbf{5}}
    & $\gamma_{1,E},\ \gamma_{2,E},\ b_{3,E},\ b_{4,E},\ p_z^{\rm rad}$ \\
  \hline
  Single Higgs @ NLO (real)
    & \textcolor{cyan}{\textbf{5}}
    & $(H\text{ width}),\ \gamma_{1,E},\ b_{3,E},\ b_{4,E},\ p_z^{\rm rad}$ \\
  \hline
\end{tabular}
}

\captionof{table}{
Summary of integration dimensions and choices of integration variables used in the MEM calculation for each LO and NLO processes.
Values in parentheses indicate the reduced dimensionality obtained
after constraining the width of off-shell particles (from Breit-Wigner propagators to
the Narrow width approximation).
}
\label{tab:integration_variables}

\end{minipage}
\end{center}

\twocolumngrid
\par\medskip

\end{document}